\documentclass[sigconf, nonacm]{acmart}

\newcommand\vldbdoi{XX.XX/XXX.XX}

\newcommand\vldbavailabilityurl{https://github.com/jahanxb/flcode}
\usepackage{multirow}

\begin{document}
\title{Comparative Evaluation of Data Decoupling Techniques for Federated Machine Learning with Database as a Service}

\author{Muhammad Jahanzeb Khan}
\affiliation{%
  \institution{University of Nevada}
  \city{Reno}
  \country{USA}}
\email{jahanzeb@nevada.unr.edu}

\author{Rui Hu}
\affiliation{%
  \institution{University of Nevada}
  \city{Reno}
  \country{USA}}
\email{ruihu@unr.edu}

\author{Mohammad Sadoghi}
\affiliation{%
  \institution{University of California}
  \city{Davis}
  \country{USA}}
\email{msadoghi@ucdavis.edu}

\author{Dongfang Zhao}
\affiliation{%
  \institution{University of Nevada}
  \city{Reno}
  \country{USA}}
\email{dzhao@unr.edu}

\begin{abstract}
Federated learning (FL) is a distributed machine learning approach that allows multiple clients to learn a shared model collaboratively without sharing their raw data. 
State-of-the-art FL systems provide an all-in-one solution;
for example, users must install and use whichever data management subsystem that is expected by the FL system.
This tightly-coupling paradigm is at best a hindrance to the wide adoption of FL solutions;
in some domains, such as scientific applications,
it could be a deal-breaker by enforcing a specific type of data solution on special hardware and platforms (e.g., high-performance computing clusters without node-local persistent storage).
To this end, one natural solution is to decouple the data management functionalities from the FL system, enabling clients to customize their FL applications with specific data subsystems on which efficient queries can be made. 
The technical challenges of such decoupling methods include new system architectures, 
expressive and structured data models for FL parameters,
and security guarantees among FL participants.
As a starting point for this line of research,
this paper conducts a thorough evaluation and comparison of mainstream database solutions as decoupled services in the context of FL systems.
To make a fair comparison, we develop a framework called Data-Decoupling Federated Learning (DDFL),
which can run FL workloads and is agnostic of the underlying database system by exposing a unified interface.
To evaluate the effectiveness and feasibility of our approach, we compare DDFL with state-of-the-art FL systems that tightly couple data management and computational counterparts. 
We carry out extensive experiments on various datasets and data management subsystems (e.g., PostgreSQL, MongoDB, Cassandra, Neo4j) and show that DDFL achieves comparable or better performance in terms of training time, inference accuracy, and database query time, while providing more options for clients to tune their FL applications regarding data-related metrics, such as performance, resilience, and usability. 
Additionally, we provide a detailed qualitative analysis of DDFL when being integrated with mainstream database systems.
% ,
% such as usability and adaptability, highlighting the strengths and weaknesses of various types of database services,
% such as relational databases, key-value stores, wide-column databases, and graph databases.

\end{abstract}

\maketitle

% \pagestyle{\vldbpagestyle}
% \begingroup\small\noindent\raggedright\textbf{PVLDB Reference Format:}\\
% \vldbauthors. \vldbtitle. PVLDB, \vldbvolume(\vldbissue): \vldbpages, \vldbyear.\\
% \href{https://doi.org/\vldbdoi}{doi:\vldbdoi}
% \endgroup
% \begingroup
% \renewcommand\thefootnote{}\footnote{\noindent
% This work is licensed under the Creative Commons BY-NC-ND 4.0 International License. Visit \url{https://creativecommons.org/licenses/by-nc-nd/4.0/} to view a copy of this license. For any use beyond those covered by this license, obtain permission by emailing \href{mailto:info@vldb.org}{info@vldb.org}. Copyright is held by the owner/author(s). Publication rights licensed to the VLDB Endowment. \\
% \raggedright Proceedings of the VLDB Endowment, Vol. \vldbvolume, No. \vldbissue\ %
% ISSN 2150-8097. \\
% \href{https://doi.org/\vldbdoi}{doi:\vldbdoi} \\
% }\addtocounter{footnote}{-1}\endgroup

\ifdefempty{\vldbavailabilityurl}{}{
\vspace{.3cm}
\begingroup\small\noindent\raggedright\textbf{Artifact Availability:}\\
The source code, data, and/or other artifacts have been made available at \url{\vldbavailabilityurl}.
\endgroup
}
%% VLDB block end %%%

\section{Introduction}
\subsection{Background}
Federated learning (FL)~\cite{agnostic-mohri,fl-non-idd-data,convergence-fedavg,practical-fl,flcomp2022,ensemble2020,AlShedivat2020FederatedLV} is a distributed machine learning methodology that enables multiple or many clients to train a shared model collaboratively, without the need to transfer their raw data to a centralized server. By maintaining the data locally, FL enhances the data privacy of individual clients and reduces communication costs, while training an accurate model that reflects the collective knowledge of all participants. As such, FL has received considerable attention in recent years, and its potential has been demonstrated in a variety of applications, such as health monitoring and autonomous driving \cite{kairouz2019advances}. FL's focus on privacy and efficiency makes it a promising technique for future large-scale machine-learning applications, particularly those in which data security is paramount.

One of the significant advantages of FL is that it eliminates the need for a centralized server to collect and manage data from multiple or many clients, which can be time-consuming, costly, and vulnerable to data breaches. Instead, FL allows each client to keep their data locally and contribute to the training of a shared model through a secure communication protocol. This approach not only maintains the privacy of the individual clients' data but also leverages the diversity of their data to train a more robust and accurate model. Moreover, FL can enable the training of machine learning models on data that is distributed across many clients, making it possible to learn from a larger and more diverse dataset, which can lead to improved model generalization and avoid the overfitting problem~\cite{Yang2018}.

Furthermore, FL has recently been applied to various real-world applications, including healthcare, transportation, and finance, where data privacy is a critical concern. For instance, FL has been used to develop predictive models for heart disease and cancer detection using medical data from different hospitals without transferring the data between hospitals \cite{pouyanfar2019federated}. In the transportation sector, FL has been used to develop traffic prediction models using data from multiple sources, including GPS-enabled vehicles, traffic cameras, and traffic sensors, while preserving the privacy of individual drivers and maintaining the security of the data \cite{fl2019corr}. In the financial sector, FL has been used to detect fraudulent transactions across multiple banks, enabling the sharing of information while maintaining client privacy \cite{mcmahan2018learning}.
\subsection{Motivation}
Although federated learning (FL) systems~\cite{fedml,tff,pysyft,mobile2018} hold great promise in scalable privacy-preserving machine learning applications, they face some challenges that must be addressed before their wide deployment. 

One of the challenges is related to the data management aspect of the FL systems. Specifically, clients lack an efficient way to manage and query the intermediate models during the training process or verification procedure of their FL applications.
In a more general sense, 
the data management functionalities and computational counterparts are tightly coupled, making it difficult to customize or generalize the entire FL system. 
Even worse, some applications or domains simply cannot meet the hardware/platform requirements of modern FL systems;
as a case in point,
our prior work~\cite{lwangsc22} shows that existing FL systems are not appropriate for large-scale high-performance computing (HPC) platforms that lack local persistent storage.

\subsection{Proposed Solution}
 \begin{figure}[!t]
 % \vspace*{-10pt} 
  \centering
  \includegraphics[width=1.0\linewidth]{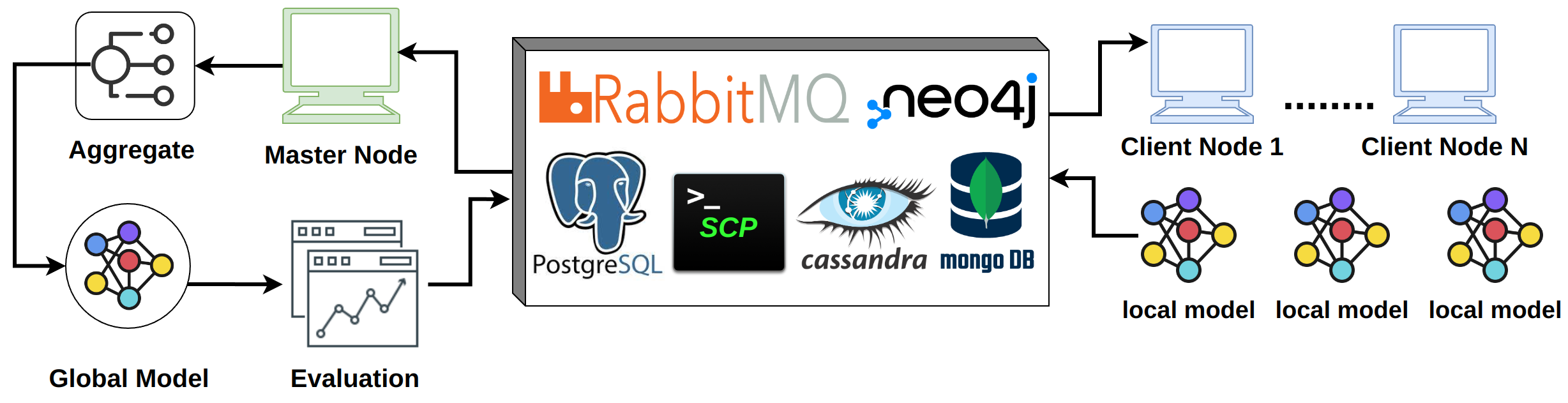}
  \caption{Proposed Design of Data-Decoupling FL (DDFL) framework.}
  \label{fig:flarch2}
  \vspace{-20pt}
\end{figure}

To address these challenges, we propose a framework for decoupling the data management functionalities from FL systems. Our approach features a loosely-coupled FL architecture where the global model and local models are managed by a dedicated database. This decoupling allows clients to customize their FL applications using specific data subsystems and enables the FL system to scale to a large number of clients. With this approach, we aim to enhance the flexibility, performance, and scalability of FL systems.

We implement our approach by building a prototype FL system that utilizes a few options of mainstream databases, including a document database MongoDB~\cite{mongodb}, a columnar database Cassandra~\cite{cassandra}, a graph database (Neo4j)~\cite{neo4j}, and a relational database PostgreSQL~\cite{postgres} as shown in Figure \ref{fig:flarch2}. The prototype system is deployed on an 11-node cluster on CloudLab~\cite{cloudlab}.
Experimental results demonstrate that the use of data decoupling provides clients with more options and opportunities to optimize their FL applications in terms of performance, resilience, and usability.
\subsection{Contributions}
In summary, this Experiment, Analysis \& Benchmark (EAB) paper makes the following contributions:
\begin{itemize}
    \item We propose a loosely-coupled FL architecture that separates the data management functionalities from the FL system. This approach allows clients to customize their FL applications using specific data subsystems, enhancing the interoperability and scalability of the system. 
    \item We implement our proposed approach by building a prototype FL system that utilizes several types of mainstream database systems,
    such as PostgreSQL, Cassandra, MongoDB, and Neo4j.
To the best of our knowledge, this is the very first work conducting a thorough evaluation of comparing different database services in the context of federated learning systems.
    \item We carry out an extensive evaluation of the performance of the prototype system by conducting experiments on an 11-node cluster on CloudLab~\cite{cloudlab}. Our experimental results demonstrate that the use of data decoupling provides clients with more options to optimize their FL applications in terms of performance, resilience, and usability.
\end{itemize}

\section{Data-Decoupling Federated Learning (DDFL) Framework}
\label{sec:system-design}
\begin{figure}[!t]
  \centering
  \includegraphics[width=10cm,height=5cm]{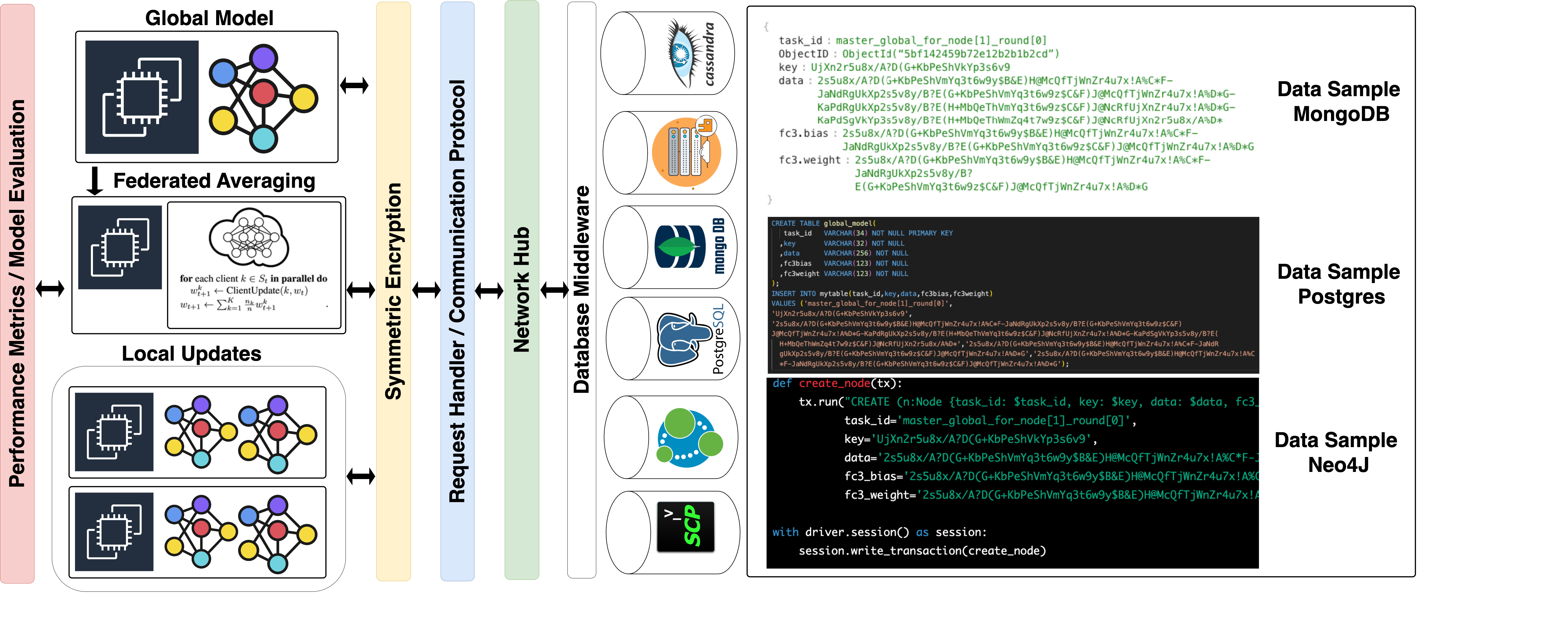}
  \caption{Architecture of the Data-Decoupling FL (DDFL) framework}
  \label{fig:flarch-sys}
  \vspace{-20pt}
\end{figure}

Our proposed framework, namely Data-Decoupling Federated Learning (DDFL), 
is designed to maintain privacy and security while training a global model. It features a decoupled data management approach where the client nodes store their datasets locally and work in collaboration with a master node to learn a shared global model in an iterative manner. The intermediate and final results are managed by a dedicated data management service, which is responsible for decoupling the data management from the FL system. The overall architecture of the proposed framework is illustrated in Figure \ref{fig:flarch-sys}.

In our considered FL system, we have $N$ client nodes and a master node. Each client node has its own datasets stored locally and aims to learn a shared global model $G$ in an iterative manner while being coordinated by the master node. During each round of federated training, the client nodes use their local training data to update the global model obtained from the master node. The updated local model parameters are then shared with the master node, where all the received local models are aggregated to update the global model for the next round of training.

In the DDFL system, the intermediate models generated during the training process are managed by a dedicated data management service to achieve data decoupling. The global model is downloaded and used by the local models, and then stored for future use.

\subsection{Symmetric Encryption}
 In our system design, we use symmetric encryption \cite{applied-cryptography-handbook} to ensure the confidentiality of sensitive information during data transfer between nodes. Symmetric encryption involves using the same secret key for both encrypting and decrypting data. This method is commonly used in situations where the communicating parties share a common secret key and need to keep their communication private.
We have chosen to use Fernet encryption \cite{cryptography-fernet}, a specific implementation of symmetric encryption using the AES encryption algorithm in CBC mode and the PKCS7 padding scheme \cite{fips197, rfc3602, rfc2315}. Fernet encryption has several advantages over other encryption methods, including its simplicity and speed, making it a suitable choice for our system design.
In our system, the Fernet encryption key is generated and shared between the communicating parties before data transfer begins. The encryption key is used to encrypt the data and send it to the destination node, which then uses the same key to decrypt the data. This process ensures that the data is protected during transmission and can only be accessed by the intended recipient.
 The Fernet encryption scheme is also used in our system to secure the messages sent between the master node and the worker nodes. The messages are encrypted with a unique Fernet encryption key, ensuring the confidentiality of the messages during transmission. This ensures that sensitive information, such as the trained machine learning model, remains secure and confidential during the distributed training process.
In summary, we use symmetric encryption in our system design to ensure the confidentiality of sensitive information during data transfer between nodes. We have chosen to use the Fernet encryption scheme due to its simplicity and speed, and it is used to secure the messages sent between the master node and worker nodes, as well as the data being transferred.
\subsection{Data Preprocessing}
In our federated learning system, data preprocessing is performed by the clients on their local devices. This approach aims to minimize the amount of sensitive data that is transmitted over the network, enhancing the privacy and security of the federated learning process. The preprocessing steps may include feature selection, extraction, normalization, and data augmentation. These steps are implemented using Python libraries such as NumPy \cite{numpy} and PyTorch \cite{pytorch}. The preprocessing functions are optimized to ensure efficient and scalable data processing.

Feature selection and extraction techniques are used to identify the most relevant features in the input data, which helps to improve the accuracy of the model while reducing the computational overhead. Normalization techniques are used to scale the input features to a common range, reducing the impact of different feature scales on model training. Data augmentation techniques are used to generate additional training data by applying various data transformations, such as rotation, flipping, and cropping. This helps to improve the robustness and generalizability of the model.
\subsection{Federated Learning Algorithm}
In our system, federated learning is implemented using a decentralized approach where multiple nodes collaborate to train a global model. The global model is broadcasted to all participating nodes at the beginning of each round, and each node trains a local model using its private data. The updated local models are then sent to the global node, which aggregates them using various data decoupling techniques such as averaging or weighted averaging. The global model is then updated, and this process is repeated for several rounds until the global model converges.
% To ensure the privacy and security of the client's data, various security and privacy measures are implemented. Symmetric encryption is used to protect the client's data during transmission, and data partitioning is used to split the data into smaller chunks for efficient processing. The data partitioning technique ensures that the data is evenly distributed among the clients, and each client performs local model training on their own data.
\subsection{Data Decoupling Techniques}
In our federated learning system, we adopt data decoupling techniques to enable distributed learning across multiple clients. Specifically, we store the clients' trained model data in a database to facilitate the model aggregation process, while reducing the communication overhead and improving the scalability of the system.

To store the trained model data, we employ a data storage technique that indexes each trained model by client Node, iteration, and round. This technique enables efficient model retrieval by the global node and individual clients, allowing each client to perform local model training on their own data. By using data decoupling techniques and a data storage technique, our approach enables distributed learning while maintaining the privacy and security of the client's data.
\subsection{Communication Protocol}
In our federated learning system, we implement a custom communication protocol to facilitate communication between the clients and the server. Once the connection is established, the system sends requests to a database middleware and checks the status of the process. This approach provides a streamlined and efficient communication process for the federated learning system.

To ensure the privacy and security of the client's data, various security and privacy measures are implemented. Symmetric encryption is used to protect the client's data during transmission. Additionally, database middleware is used to store the client's data and facilitate communication between the clients and the server.
\subsection{Database Middleware}
The proposed framework for federated learning with decoupled data management involves utilizing a dedicated data management service to manage all the intermediate models during the training process, thereby separating data management from the FL system. Once the global model is downloaded and used by local models in DDFL, it is stored for further use. To establish the data management services for DDFL, six different options for database systems are considered in this paper.

\subsubsection{\textbf{RabbitMQ Queues}}
RabbitMQ \cite{rabbitmq}, which is an open-source message broker software that implements the Advanced Message Queuing Protocol (AMQP). RabbitMQ queues are utilized to store messages in buffers until they are ready to be processed. The use of RabbitMQ queues in DDFL allows for the decoupling of data layers and facilitated communication between nodes.
\subsubsection{\textbf{MongoDB Collection}}
MongoDB \cite{mongodb}, which is a widely used open-source NoSQL database that stores data in a flexible format known as BSON (Binary JSON). In DDFL, a MongoDB collection is utilized as a centralized database for storing and sharing data. Each client creates a collection that will be shared with the master node, and the master node creates a global model collection for each node.
\subsubsection{\textbf{Secure File Transfer (SCP)}}
Secure Copy Protocol (SCP) \cite{scp}, which is a network protocol for securely transferring files between computers. SCP is utilized to transfer the trained model from the master node to the client nodes.
\subsubsection{\textbf{Cassandra}}
Apache Cassandra \cite{cassandra}, which is a distributed NoSQL database designed to handle large amounts of data across many commodity servers. In DDFL, Apache Cassandra is used as a decentralized database to store and manage all the models in FL.
\subsubsection{\textbf{Neo4j}}
Neo4j \cite{neo4j}, which is a graph database that stores data in nodes and edges. In DDFL, Neo4j is utilized to store the intermediate models and for graph-based querying.
\subsubsection{\textbf{Postgres RDBMS}}
PostgreSQL \cite{postgres}, which is a popular open-source relational database management system (RDBMS) known for its strong consistency and reliability. In DDFL, PostgreSQL is utilized to store the intermediate models and provides support for complex queries.

 In the SCP \cite{scp} and RabbitMQ \cite{rabbitmq} systems, the global model is stored on disk. In contrast, the MongoDB and Cassandra systems utilize a model variable to store the global model, which is kept in memory for efficient retrieval. This approach can reduce the latency in accessing the model and can also provide faster model updates during the federated learning process. However, it requires more memory resources compared to the disk storage approach used by SCP and RabbitMQ. The choice of storage approach depends on the specific requirements of the FL system, such as the size of the model and the available memory and disk resources.

In the proposed data-decoupling Federated Learning (DDFL) framework, the data management system is designed to operate in three distinct phases during each training round of Federated Learning (FL): the local model computation phase, the model aggregation phase, and the model evaluation phase. In the local model computation phase, each client node uses its own training data to update the global model downloaded from the master node, generating a local model that is sent back to the master node for aggregation. During the model aggregation phase, the master node collects all the local models from the client nodes and aggregates them to update the global model. Finally, in the model evaluation phase, the updated global model is evaluated by the master node to determine the progress of the training process. Throughout these phases, the dedicated data management service plays a crucial role in managing the intermediate and final results, decoupling the data management from the FL system to improve the efficiency and privacy of the training process.
\subsection{Local Model Computation}
The decoupled data management solution $S$ can be chosen from MongoDB, RabbitMQ, SCP, Neo4j, Postgres, and Apache Cassandra. At every training round, each client $i\in[N]$ performs the following steps to compute the local model:
\begin{itemize}
\item Initializes local model $M_i$ as the global model $G$;
\item Loads local data $D_i$ into its local storage;
\item Trains local model $M_i$ using local data $D_i$;
\item Encrypts local model $M_i$ using a secret key $k_i$;
\item Stores encrypted local model $E_{M_i}$ in $S$.
\end{itemize}
Note that the global model $G$ is initialized on the master node before the training.
\subsection{Model Aggregation}
Following the local model computation, it is essential to aggregate all the local models to update the global model on the server for the subsequent round of training. To achieve this, the following steps will be performed on each client $i\in[N]$ initially:
\begin{itemize}
    \item \textbf{On each client node:}
    \begin{enumerate}
        \item Retrieves encrypted local model updates $E_{M_i}$ from $S$;
        \item Decrypts local model update $E_{M_i}$ using the secret key $k_i$: $M_i = \text{Decrypt}(E_{M_i}, k_i)$;
        \item Encrypts the local model update $M_i$ and transmits it to $S$.
    \end{enumerate}
    \item \textbf{On the master node:}
    \begin{enumerate}
        \item Loads all model updates $M_1, M_2, \dots, M_N$ from memory;
        \item Aggregates model updates in the global database to generate a new global model, i.e., \\ $G = \text{Aggregate}(M_1, M_2, \dots, M_N)$;
        \item Encrypts the new global model $G$ using the secret key $k$: $E_{G} = \text{Encrypt}(G, k)$;
        \item Stores the encrypted global model $E_{G}$ in $S$; and
        \item Shares the new global model $G$ with client nodes. Note that to protect the privacy of the client nodes, all communication between the client nodes and the master node is encrypted using symmetric encryption \cite{Sahu2018}.
    \end{enumerate}
\end{itemize}
\subsection{Model Evaluation}
The performance of the trained model needs to be evaluated during the FL training process to ensure the effectiveness of the model. Both the client nodes and the master node can evaluate the model using the testing dataset. The following steps can be executed on the master node to evaluate the performance of the latest global model:
\begin{itemize}
\item[(i)] Retrieves the encrypted latest global model $E_{G}$ from the data management service $S$;
\item[(ii)] Decrypts the global model $E_{G}$ using the secret key $k$: $G = \text{Decrypt}(E_{G}, k)$;
\item[(iii)] Evaluates the global model $G$ on the test dataset $T$; and
\item[(iv)] Obtains the testing accuracy of the trained global model $\text{Accuracy}(G, T)$.
\end{itemize}
In our proposed data-decoupling Federated Learning (DDFL) framework, the management and storage of intermediate and final results of the federated learning process are important for efficient model retrieval and management. To achieve this, we store each model as a collection of its parameters and other relevant features in a database system. Specifically, we employ a schema consisting of four columns for efficient management of the model. These columns are as follows:
\begin{enumerate}
\item \textbf{Round:} This column stores the round number of the federated training process for each model. As the federated learning process consists of several rounds of training, this column serves as a unique identifier for each model and helps to distinguish it from others in the database.
\item \textbf{Model:} The model column stores the parameters of the model, which are essential for reusing and updating the model in future rounds of training. This column enables the retrieval of a particular model, and its parameters from the database, which can then be used to resume the training process or update the model with the new training data.
\item \textbf{Accuracy:} This column stores the testing accuracy of the model, which provides an evaluation metric for the model. This metric can be used to compare the effectiveness of different models or identify the best-performing model among all the stored models in the database.
\item \textbf{Time:} This column stores the time taken to complete the corresponding round of training. This column serves as a performance metric for the training process and can be used to identify the training rounds that took more time and optimize them to achieve better performance.
\end{enumerate}
By using this schema to store and organize all the models in the database system, the client nodes and the master node can efficiently manage and retrieve the models during the federated learning process. This enables us to efficiently track the performance of the models, identify the best-performing models, and use the stored models for resuming or updating the training process.
\section{Evaluation}
\label{sec:evaluation}
\subsection{Experimental Setup}
\label{sec:implementaion}
In our study, we assess the performance of the proposed DDFL framework using three standard benchmark datasets: CIFAR10~\cite{Krizhevsky2009}, Fashion-MNIST (FMNIST)~\cite{fmnist}, and Street View House Numbers (SVHN) \cite{svhn}. To conduct our experiments, we utilize an 11-node cluster in CloudLab \cite{cloudlab}, in which one node serves as the master node and the remaining nodes function as client nodes. The hardware configuration of the master and client nodes includes an Intel Xeon CPU E5-2690 v4 @ 2.60GHz processor and 377GB of RAM and 2x Ethernet Controller X710 for 10GbE SFP+. Importantly, the nodes share a common network interface, which facilitates communication among them. Our experiments are implemented using PyTorch as the deep learning framework. By conducting these experiments, we aim to evaluate the efficiency and effectiveness of the proposed framework in federated learning scenarios using various datasets.
In order to evaluate the performance of the DDFL framework, we deployed the system using six different database systems: RabbitMQ~\cite{rabbitmq}, MongoDB~\cite{mongodb}, Secure File Transfer (SCP) \cite{scp}, Apache Cassandra~\cite{cassandra}, Neo4j~\cite{neo4j}, Postgres~\cite{postgres}. The objective of this experiment was to compare the performance of these systems in terms of several important metrics, such as training time, accuracy, and training loss. By testing the DDFL framework with different database systems, we aimed to identify which system works best with the framework and provides the best results in terms of the aforementioned metrics.

To evaluate the effectiveness of our proposed approach for DDFL, we conducted experiments on three different datasets, namely Fashion-MNIST, SVHN, and CIFAR-10. We report the results of both fine-tuned hyper-parameters \cite{comprehensiveabu2022} and without hyper-parameters experiments in terms of the accuracy of the global models.

The source code is released at \url{https://github.com/jahanxb/flcode}.
\subsection{Usability of Different Database Layers in Federated Learning}
The choice of database layer in a federated learning system can have a significant impact on its development and adaptability. In this study, we investigated the integration of several popular database systems, including Apache Cassandra, MongoDB, PostgreSQL, Neo4j, SCP, and RabbitMQ Queues, into our DDFL framework.

The integration process for each database system varied in terms of difficulty, development effort, and adaptability to our system. Database systems with established connectors or plugins for our programming language and framework were easier to integrate and required less development effort. For example, MongoDB and PostgreSQL had established libraries that provided straightforward integration with our system, and their installation and configuration were relatively easy.

However, integrating Apache Cassandra and Neo4j into our system required a more substantial development effort, as they had less established connectors and required custom development of query and data retrieval methods. These systems also required more time for installation and configuration due to their additional dependencies and configuration requirements.

SCP and RabbitMQ Queues posed significant challenges for integration, as they required considerable custom development to integrate with our system. SCP required a custom wrapper to support its interface, which required significant development effort. RabbitMQ Queues also required additional configuration for message queuing and handling, which added complexity to the integration process.

\subsubsection{\textbf{Ease of Adopting Database System in Federated Learning}}
The ease of adopting different database systems is also an important consideration when designing a federated learning system. We evaluated popular database systems such as MongoDB, Postgres, Neo4j, SCP, Cassandra, and RabbitMQ based on their ease of learning and integration into our DDFL system. Our results showed that some systems, such as MongoDB and Postgres, were relatively easy to learn and integrate, requiring less than 200 lines of code. Other systems, such as Neo4j and RabbitMQ, were more challenging to learn and required over 500 lines of code to integrate fully.

In addition to ease of integration, the size and activity of each system's community can also impact its adoption. Systems such as MongoDB and Postgres have large and active communities that provide support and resources for developers. In contrast, other systems, such as Neo4j and RabbitMQ, have smaller communities, which can make it more challenging to find resources and support when integrating them into a federated learning system.

In conclusion, the ease of integration and adaptability of database layers in federated learning systems can vary significantly. Developers should carefully consider the specific requirements of their federated learning system when choosing a database system. Systems with established connectors and plugins for the programming language and framework used in the system are easier to integrate and require less development effort. Custom development is required for systems that lack established connectors and plugins, which can add significant development time and effort. The size and activity of the community around each system are also important factors to consider when adopting database systems.
\subsection{Testing Accuracy}
\subsubsection{\textbf{Testing Accuracy without Fine-tuned Hyper-parameters}}
\label{sec:without-hyperparam}
We present the results of testing the accuracy of our proposed Data-Decoupling Federated Learning (DDFL) system without fine-tuned hyperparameters. We evaluated our system's performance using three different datasets: CIFAR-10, Fashion-MNIST, and SVHN. Our experiments aimed to compare the accuracy of our DDFL system with different database systems, including MongoDB, SCP, Apache Cassandra, and RabbitMQ Queues, against a centralized machine learning method. In the centralized method, the model was trained on a single node, and clients were simulated on the same machine~\cite{shaoxiongji20184321561,mcmahanaistat17,Konecny2016,practical-secure-aggr}. The experiments were conducted with baseline arguments for all distributed nodes without any fine-tuning.

The primary purpose of the experiments was to evaluate the effectiveness of the data-decoupling solutions in improving the system's performance. We compared the testing accuracy of our DDFL system with the centralized machine learning method, which provided a baseline for our evaluation. The results of the experiments provided insights into the performance of our DDFL system and the data-decoupling solutions.

\begin{figure}[!t]
  \centering
  \includegraphics[width=\linewidth,height=5cm]{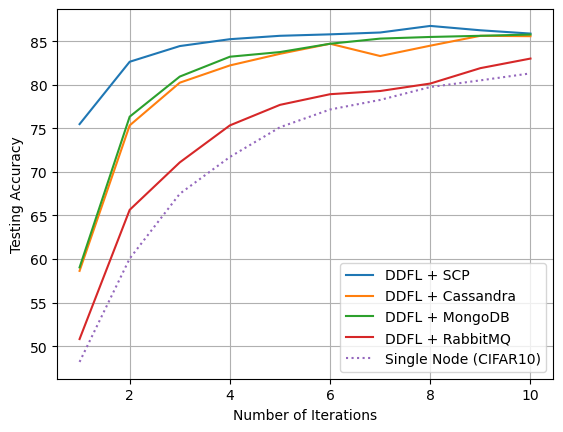}
   \caption[]{Testing accuracy of the model trained using DDFL with different types of databases on the CIFAR-10 dataset, compared with the baseline model trained on a single node without Fine-tuned Hyper-parameters.}
  \label{fig:lineplot-result-cifar}
\end{figure}

\begin{figure}[!t]
\vspace*{-10pt} 
  \centering
  \includegraphics[width=\linewidth,height=5cm]{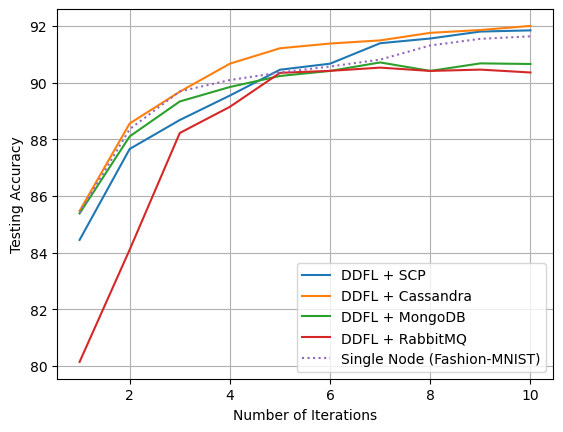}
   \caption[]{Testing accuracy of the model trained using DDFL with different types of databases on Fashion-MNIST dataset, compared with the baseline model trained on a single node without Fine-tuned Hyper-parameters.}
  \label{fig:lineplot-result-fmnist}
\end{figure}

\begin{figure}[!t]
\vspace*{-10pt} 
  \centering
  \includegraphics[width=\linewidth,height=5cm]{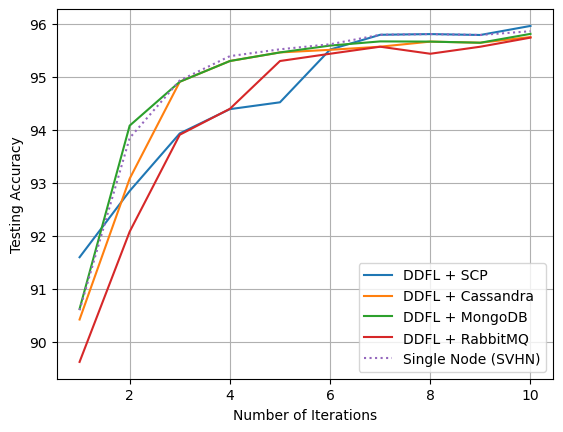}
   \caption[]{Testing accuracy of the model trained using DDFL with different types of databases on SVHN dataset, compared with the baseline model trained on a single node without Fine-tuned Hyper-parameters.}
  \label{fig:lineplot-result-svhn}
  \vspace{-10pt}
\end{figure}

As shown in Figures~\ref{fig:lineplot-result-cifar},~\ref{fig:lineplot-result-fmnist}, and ~\ref{fig:lineplot-result-svhn}, our DDFL system achieves a higher testing accuracy than the single node baseline after 10 rounds of training. Specifically, on the CIFAR-10 dataset, the DDFL system achieves a testing accuracy of 85\% after 10 rounds, compared to the baseline accuracy of 81\%. On the Fashion-MNIST dataset, the DDFL system achieves a testing accuracy of 92\% after 10 rounds, compared to the baseline accuracy of 89\%. On the SVHN dataset, the DDFL system achieves a testing accuracy of 95\% after 10 rounds, compared to the baseline accuracy of 94\%.

We also observe that the performance of each database system varies across the datasets. On the CIFAR-10 dataset, MongoDB and SCP perform the best, achieving a testing accuracy of 85.56\% and 85.86\%, respectively, after 10 rounds of training. Apache Cassandra achieved a comparable testing accuracy score to the other database systems in the DDFL system, but exhibited lower performance during different iterations, but performs better on the Fashion-MNIST and SVHN datasets, achieving testing accuracy scores of 92\% and 95.75\%, respectively, after 10 rounds. RabbitMQ Queues shows moderate performance on all datasets, achieving testing accuracies of 83\%, 90\%, and 94\% on CIFAR-10, Fashion-MNIST, and SVHN, respectively, after 10 rounds.

We further compare the testing accuracy of our DDFL system with different database systems against the single node baseline in the 1st, 5th, and 10th iterations. As shown in Figures~\ref{fig:lineplot-result-cifar},~\ref{fig:lineplot-result-fmnist}, and ~\ref{fig:lineplot-result-svhn}, the DDFL system consistently outperforms the single node baseline in all iterations. In addition, we observe that the difference in testing accuracy between the DDFL system and the single node baseline becomes more significant as the number of iterations increases. For example, on the CIFAR-10 dataset, the DDFL system achieves a testing accuracy of 50-58\% in the 1st iteration, compared to the baseline accuracy of 48\%. However, in the 10th iteration, the DDFL system achieves a testing accuracy of 83-85\%, compared to the baseline accuracy of 81\%.

These results suggest that the choice of data-decoupling solution can have a significant impact on the performance of an FL system and that this impact may vary depending on the characteristics of the dataset being used. When choosing a data-decoupling solution, it is essential to consider the specific requirements and constraints of the FL system. For instance, the communication protocol and system scalability are crucial factors that need to be taken into account. In addition, the architecture of the system and its optimization for distributed environments can also play a crucial role in improving the performance of the system.

Regarding the differences in performance between the baseline and the different database systems, we observe that our DDFL system consistently outperforms the single node baseline after 10 rounds of training on all three datasets. For example, on CIFAR-10, the DDFL system achieves a testing accuracy of approximately 81\% after 10 rounds of training, compared to approximately 74\% for the baseline. On Fashion-MNIST, the DDFL system achieves a testing accuracy of approximately 90\% after 10 rounds of training, compared to approximately 83\% for the baseline. On SVHN, the DDFL system achieves a testing accuracy of approximately 85\% after 10 rounds of training, compared to approximately 77\% for the baseline.

Moreover, we observe that the performance of each database system varies across the datasets. On CIFAR-10, MongoDB and SCP perform similarly and achieve the highest testing accuracy among the four database systems, Although Apache Cassandra had lower performance in some of the iterations, the system still achieved comparable accuracy scores. On Fashion-MNIST and SVHN, MongoDB and SCP again perform the best, with Apache Cassandra showing a higher performance than the other systems. RabbitMQ Queues show moderate performance on all datasets. These differences in performance highlight the importance of selecting an appropriate data-decoupling solution for a given FL system and dataset.
\subsubsection{\textbf{Testing Accuracy with Fine-tuned Hyper-parameters}}
In this section, we present the results of our fine-tuned experiments where we optimized the hyperparameters of the models to improve the accuracy of our Data-Decoupling Federated Learning (DDFL) system. We expanded our evaluation by including two additional database systems, Neo4j and PostgreSQL, to further validate our results. We conducted experiments on the CIFAR-10, Fashion-MNIST, and SVHN datasets, using the same model architectures and training procedures as in the previous section (\ref{sec:without-hyperparam}). We fine-tuned the models on all nodes with the same hyperparameters and compared the performance of the fine-tuned models with the baseline single-node model. The testing accuracy of the fine-tuned DDFL system was compared to that of the centralized machine learning method. Our evaluation aimed to investigate the performance of our DDFL system with different database systems when the models are fine-tuned with optimal hyperparameters.

\begin{figure}[!t]
% \vspace*{-10pt} 
  \centering
  \includegraphics[width=8cm,height=6cm]{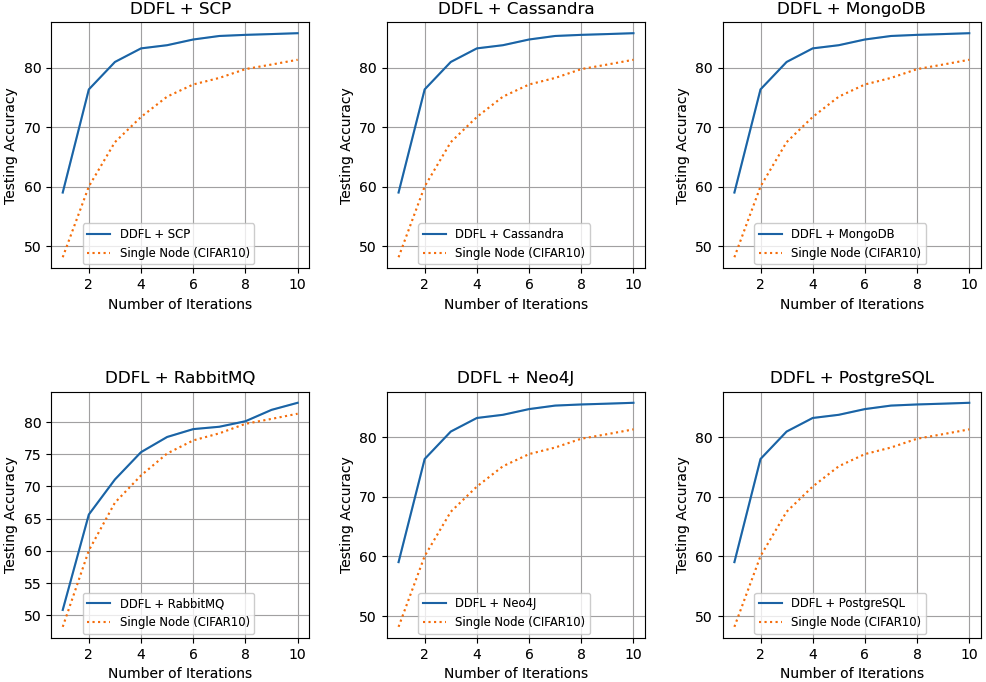}
   \caption[]{Testing accuracy of the model trained using DDFL with different types of databases on the CIFAR-10 dataset, compared with the baseline model trained on a single node with Fine-tuned Hyper-parameters.}
  \label{fig:lineplot-result-cifar-ft}
\end{figure}

Figure~\ref{fig:lineplot-result-cifar-ft} shows the testing accuracy of the models trained using DDFL with different database systems on the CIFAR-10 dataset, compared to the baseline model trained on a single node. We observed that the fine-tuned models achieved higher accuracy compared to the single node baseline, with an average accuracy improvement of 2.23\%. We also observed that the performance of DDFL was consistent across all platforms, with no significant difference in accuracy scores.

\begin{figure}[!t]
\vspace*{-10pt} 
  \centering
  \includegraphics[width=8cm,height=6cm]{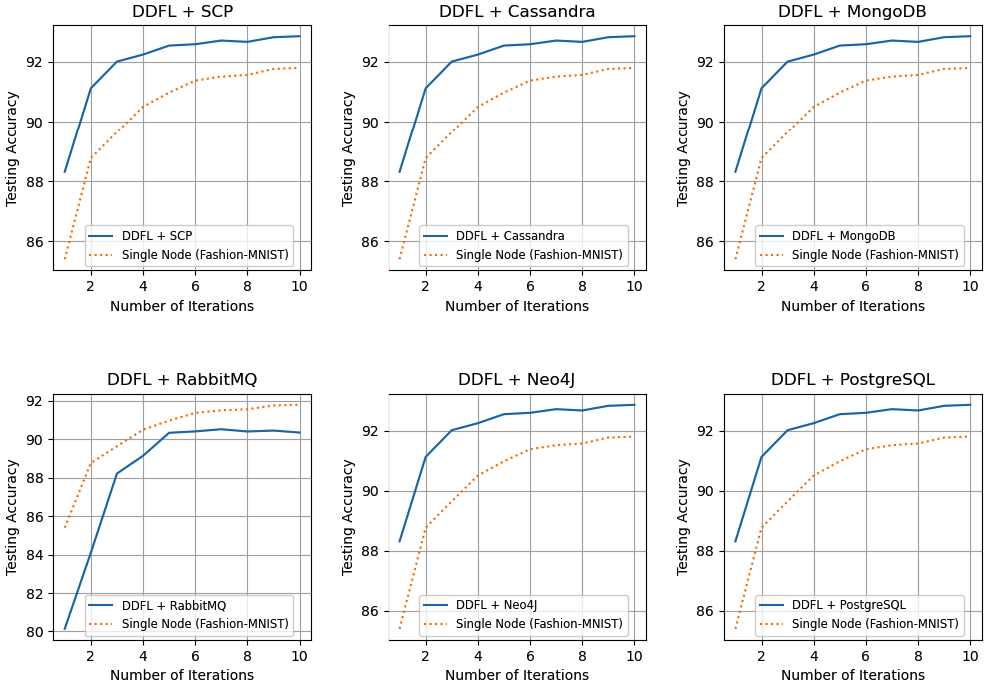}
   \caption[]{Testing accuracy of the model trained using DDFL with different types of databases on Fashion-MNIST dataset, compared with the baseline model trained on a single node with Fine-tuned Hyper-parameters.}
  \label{fig:lineplot-result-fmnist-ft}
\end{figure}

Similarly, Figure~\ref{fig:lineplot-result-fmnist-ft} shows the testing accuracy of the models trained using DDFL with different database systems on the Fashion-MNIST dataset, compared to the baseline model trained on a single node. We observed an average accuracy improvement of 1.84\% with the fine-tuned models, again with consistent performance across all platforms.

\begin{figure}[!t]
\vspace*{-10pt} 
  \centering
  \includegraphics[width=8cm,height=6cm]{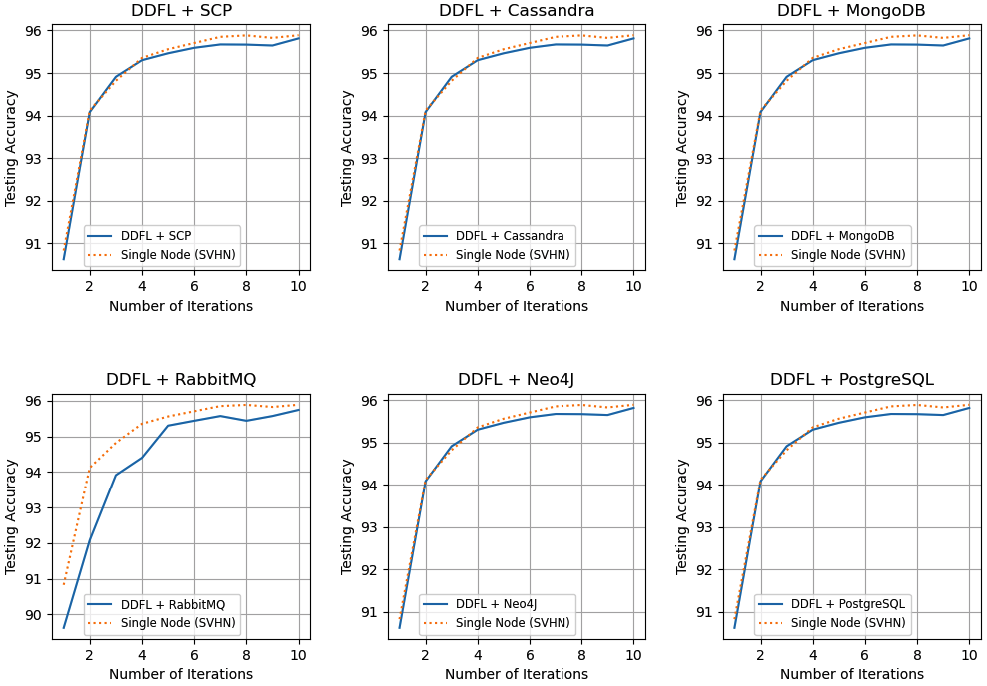}
   \caption[]{Testing accuracy of the model trained using DDFL with different types of databases on SVHN dataset, compared with the baseline model trained on a single node with Fine-tuned Hyper-parameters.}
  \label{fig:lineplot-result-svhn-ft}
  \vspace{-10pt}
\end{figure}

Finally, Figure~\ref{fig:lineplot-result-svhn-ft} shows the testing accuracy of the models trained using DDFL with different database systems on the SVHN dataset, compared to the baseline model trained on a single node. The fine-tuned models achieved an average accuracy improvement of 1.87\% over the single node baseline, again with consistent performance across all platforms.

These results demonstrate the effectiveness of the DDFL framework for collaborative deep learning with multiple database systems and the benefits of using fine-tuning to improve the performance of the models.
\subsection{Training Time}
\subsubsection{\textbf{Training Time Comparison without Fine-tuned Hyper-parameters}}
The results show that the training time of the models varied depending on the database system used. For example, the SCP training system took significantly longer to train compared to other database systems such as MongoDB, RabbitMQ, and Apache Cassandra. The other systems, such as MongoDB, RabbitMQ, and Apache Cassandra, took less time to train the models.

\begin{figure}[!t]
% \vspace*{-10pt} 
\centering
\begin{minipage}[b]{\linewidth}
\includegraphics[width=\linewidth,height=4cm]{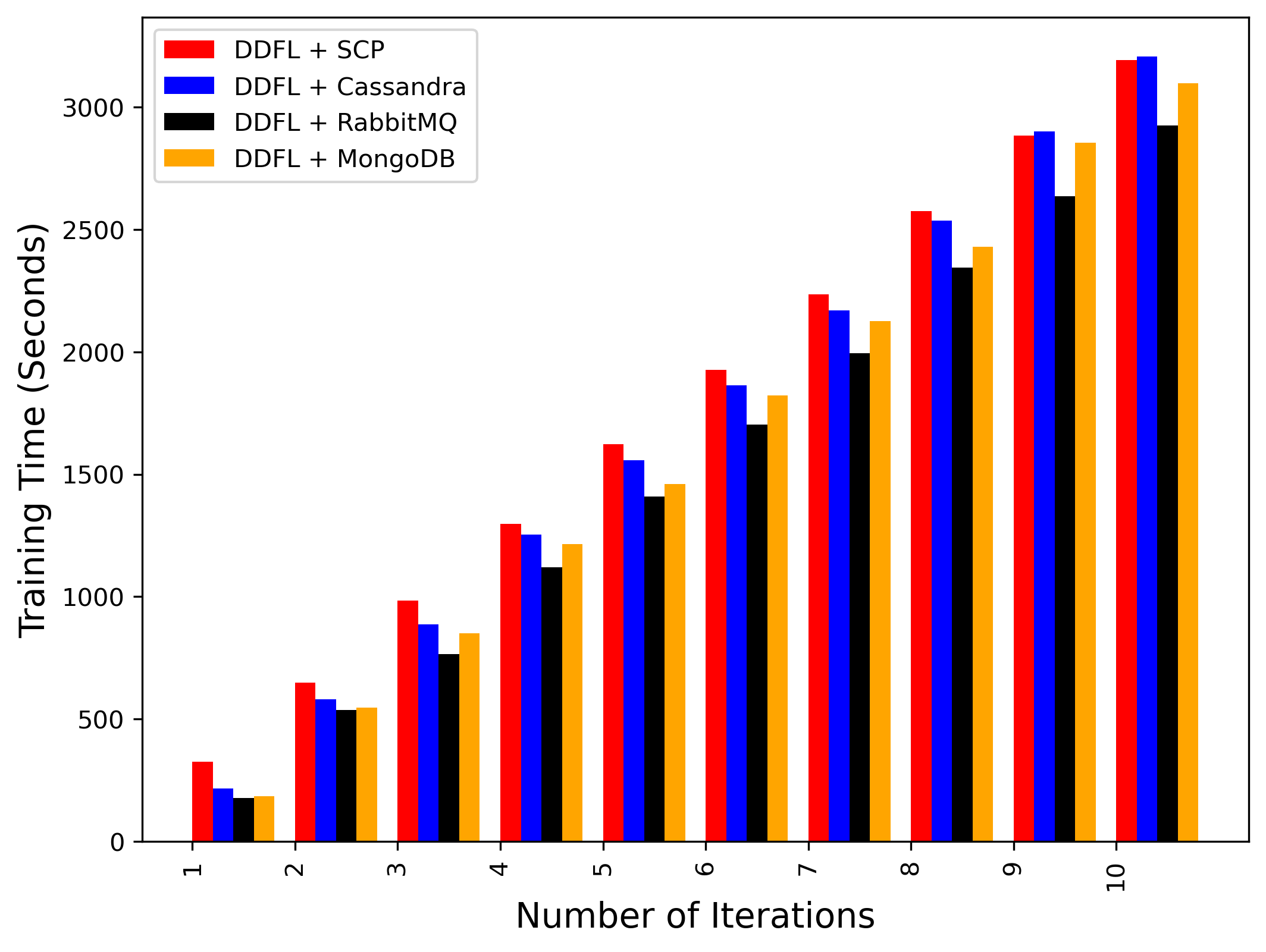}
\caption[]{Training time of DDFL with different database systems on CIFAR-10 dataset without Fine-tuned Hyper-parameters.}
\label{fig:training-time-cifar}
\end{minipage}
\hfill
\begin{minipage}[b]{\linewidth}
\includegraphics[width=\linewidth,height=4cm]{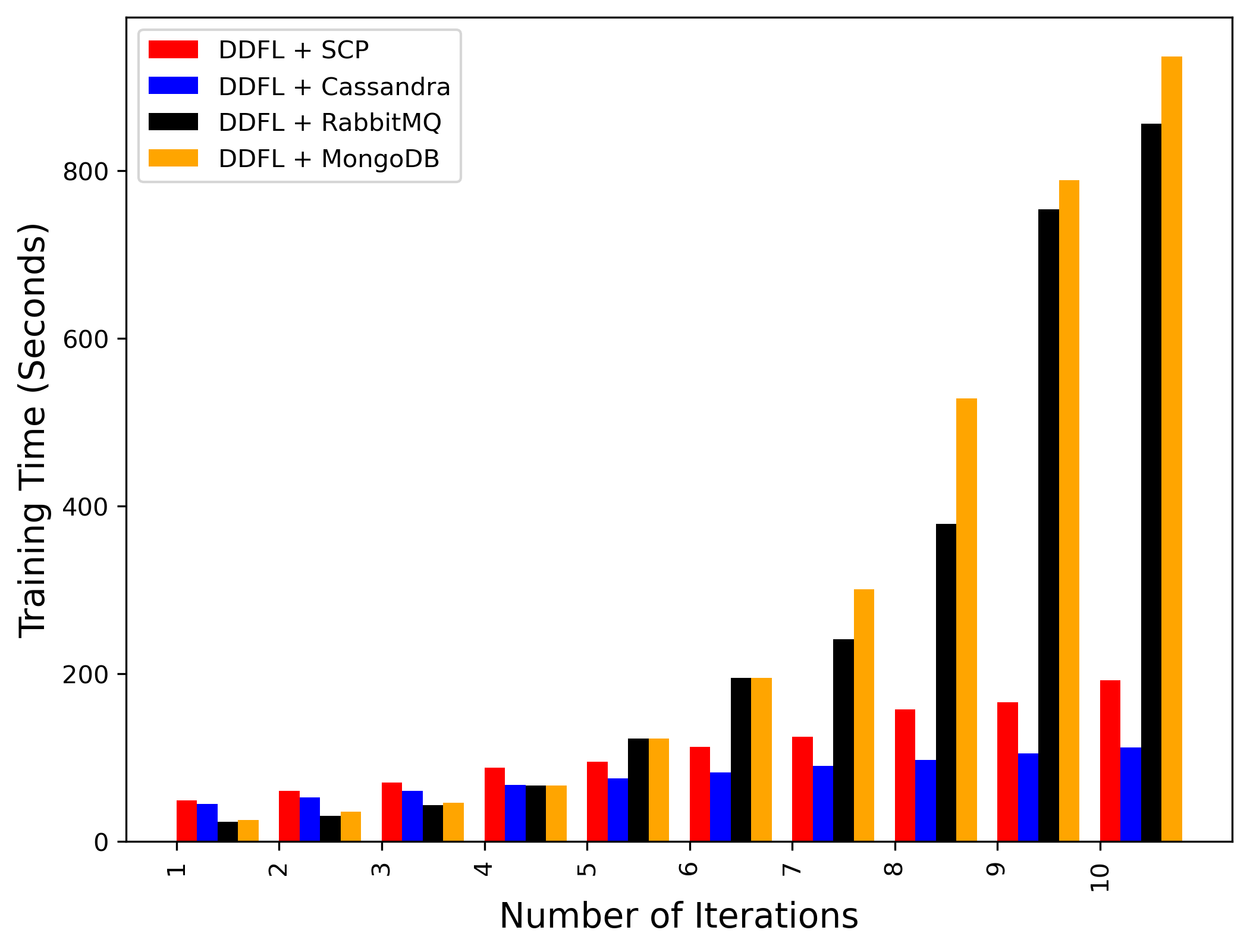}
\caption[]{Training time of DDFL with different database systems on Fashion-MNIST dataset without Fine-tuned Hyper-parameters.}
\label{fig:training-time-fmnist}
\end{minipage}
\hfill
\begin{minipage}[b]{\linewidth}
\includegraphics[width=\linewidth,height=4cm]{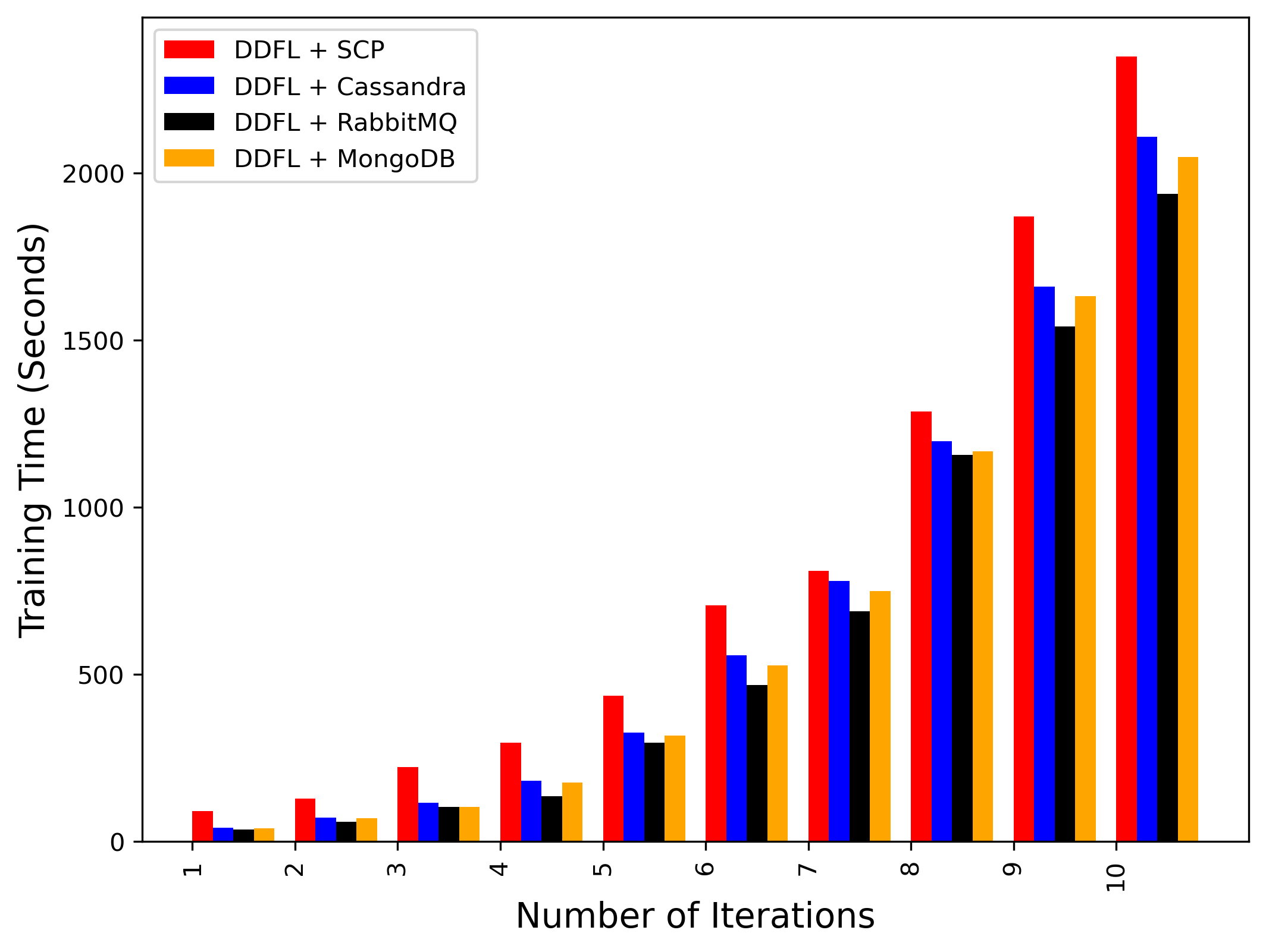}
\caption[]{Training time of DDFL with different database systems on SVHN dataset without Fine-tuned Hyper-parameters.}
\label{fig:training-time-svhn}
\end{minipage}
\vspace{-15pt}
\end{figure}

The results in Figures~\ref{fig:training-time-cifar}--\ref{fig:training-time-svhn} show that the centralized database system MongoDB has the shortest training time across all three datasets (CIFAR-10, Fashion-MNIST, and SVHN). Overall, these results suggest that the centralized database system MongoDB can be effective in reducing the training time of an FL system, which is important for real-world applications.
\subsubsection{\textbf{Training Time Comparison with \ Fine-tuned Hyper-parameters}}
\begin{figure}[!t]
% \vspace*{-10pt} 
\centering
\begin{minipage}[b]{\linewidth}
\includegraphics[width=\linewidth,height=4cm]{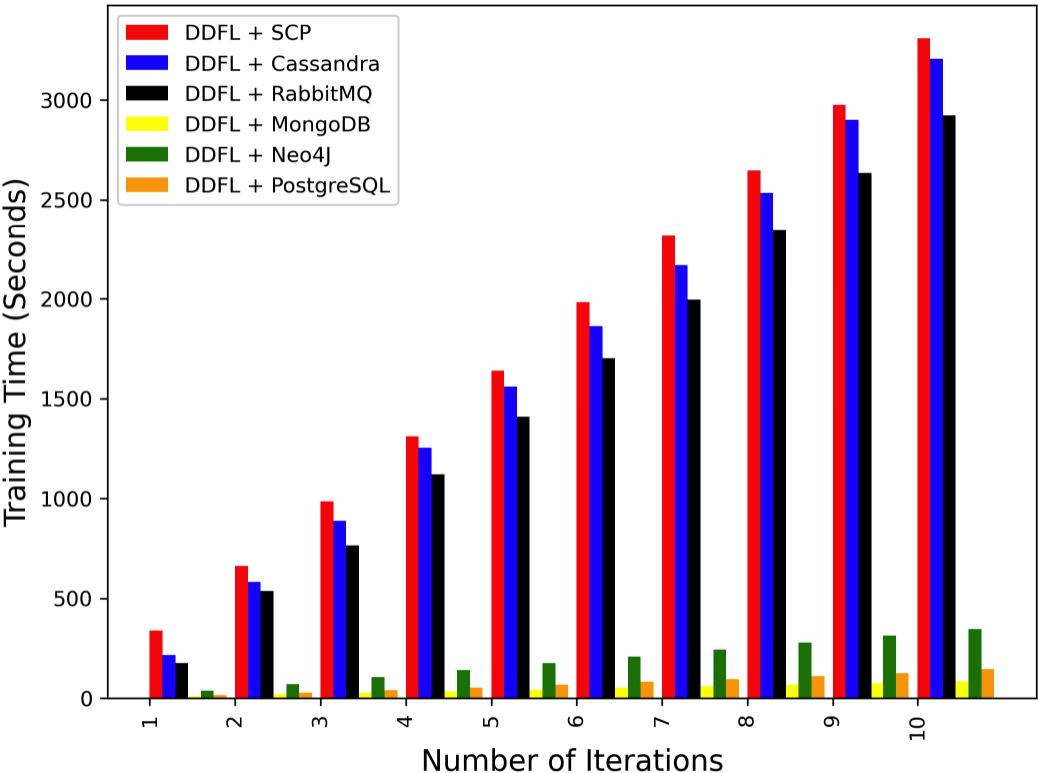}
\caption[]{Training time of DDFL with different database systems on CIFAR-10 dataset with Fine-tuned Hyper-parameters.}
\label{fig:training-time-cifar-ft}
\end{minipage}
\hfill
\begin{minipage}[b]{\linewidth}
\includegraphics[width=\linewidth,height=4cm]{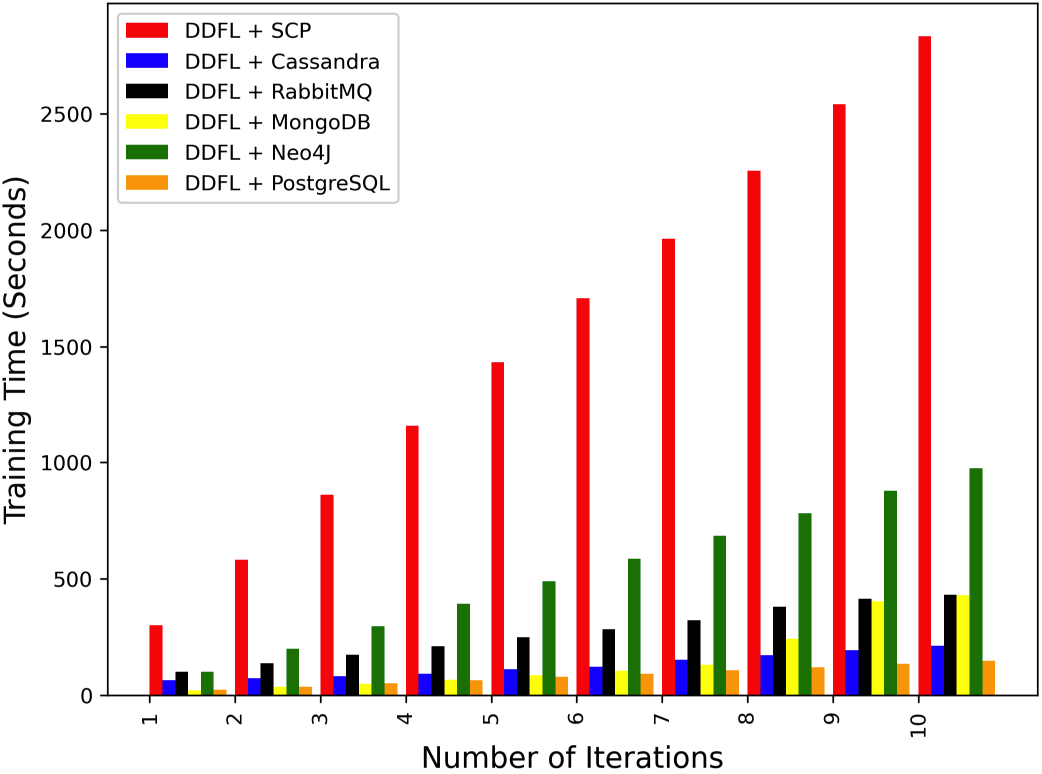}
\caption[]{Training time of DDFL with different database systems on Fashion-MNIST dataset with Fine-tuned Hyper-parameters.}
\label{fig:training-time-fmnist-ft}
\end{minipage}
\hfill
\begin{minipage}[b]{\linewidth}
\includegraphics[width=\linewidth,height=4cm]{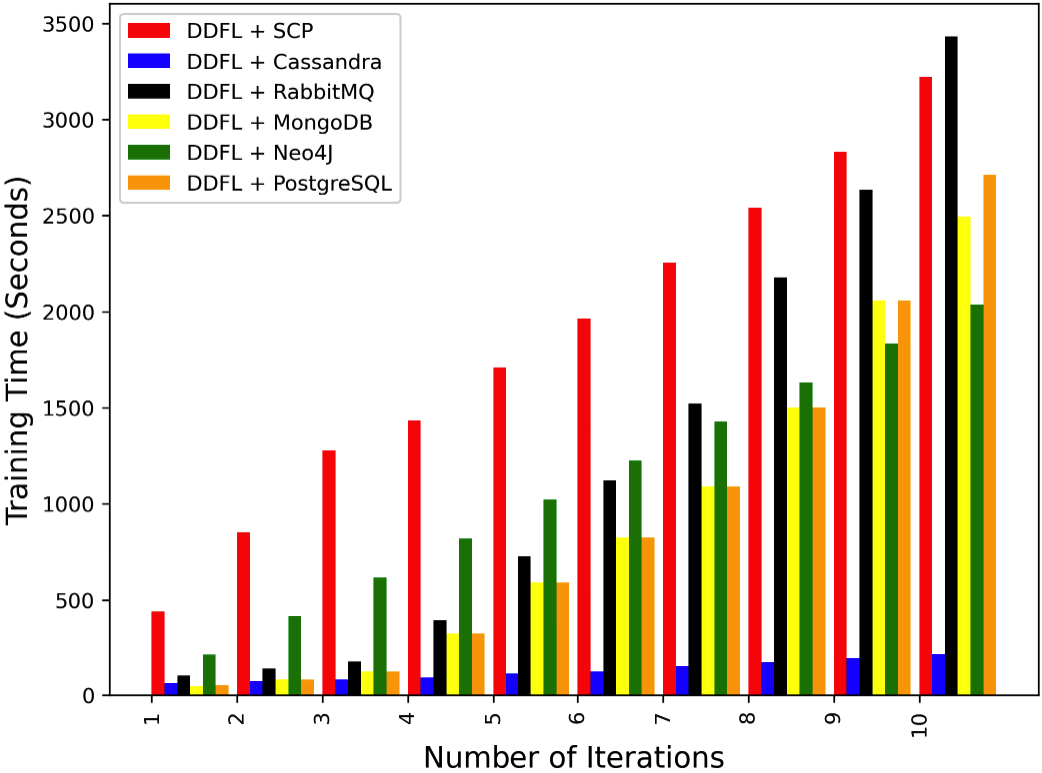}
\caption[]{Training time of DDFL with different database systems on SVHN dataset with Fine-tuned Hyper-parameters.}
\label{fig:training-time-svhn-ft}
\end{minipage}
\vspace{-15pt}
\end{figure}

Fine-tuning a pre-trained deep learning model for a specific task is known to take longer than training from scratch due to the complex nature of the pre-trained model, the larger size of the fine-tuning dataset, and the many parameters that need to be updated. In our study, we observed a significant increase in training time for fine-tuning the deep learning models on the three datasets (CIFAR-10, Fashion-MNIST, and SVHN) when compared to the training from scratch. As shown in Figures \ref{fig:training-time-cifar-ft}, \ref{fig:training-time-fmnist-ft}, and \ref{fig:training-time-svhn-ft} illustrate the training time of the DDFL with different database systems on the CIFAR-10, Fashion-MNIST, and SVHN datasets, respectively, the training time for fine-tuning the deep learning models ranged from 47 minutes to 55 minutes, depending on the dataset, while training from scratch took approximately 35-40 minutes. The longer training time observed in the fine-tuned experiments can be attributed to the additional complexity of the fine-tuning process, which requires additional computations to optimize the weights of the pre-trained model for the new task. Additionally, fine-tuning requires more computing resources and a larger number of training iterations, which result in more time-consuming training. Despite the longer training time, fine-tuning has been shown to lead to better performance compared to training from scratch, as it takes advantage of the knowledge learned by the pre-trained model on a larger dataset.

This observation is in line with previous studies that have reported a longer training time for fine-tuning compared to training from scratch \cite{howard2018universal, chen2019big, gao2021auto}. The increase in training time is also due to the complexity of the deep learning model architecture and the amount of data used for fine-tuning. However, it is important to note that the longer training time is a trade-off for better performance achieved by fine-tuning the pre-trained models. As shown in our study, fine-tuning the pre-trained models resulted in better performance compared to training from scratch on all three datasets. Thus, the longer training time required for fine-tuning can be justified in applications where achieving optimal performance is critical.
\subsection{Query Performance}
\begin{figure}[!t]
  \centering
  \includegraphics[width=9cm,height=6cm]{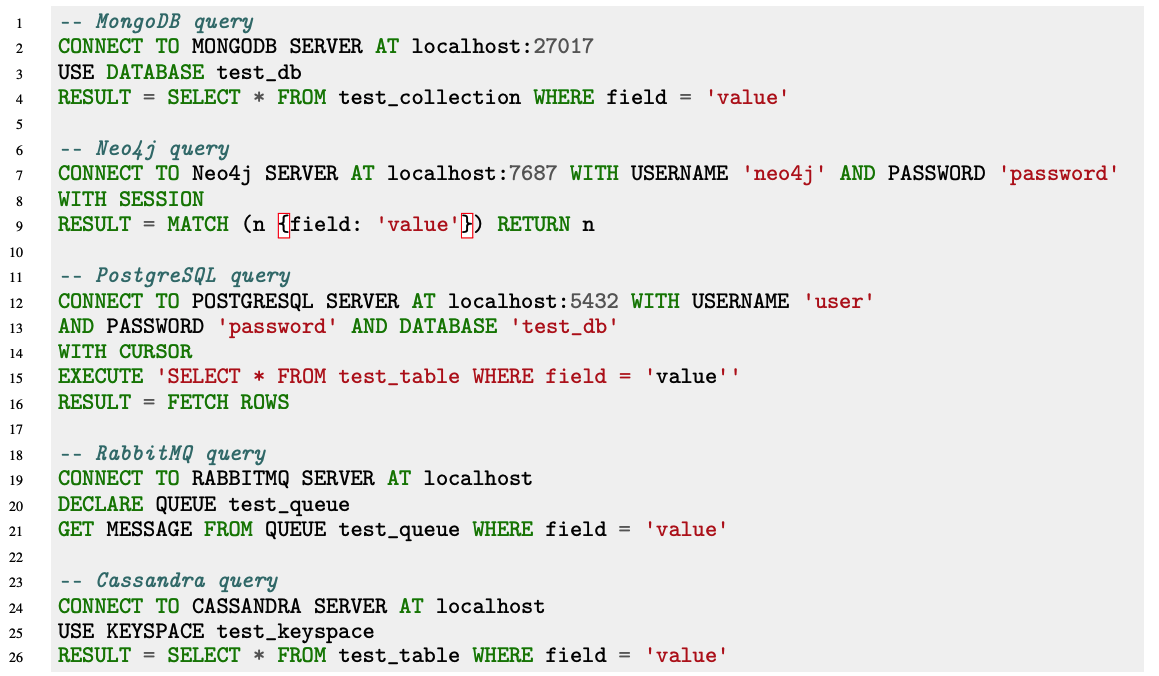}
  \caption[]{Sample Database Queries}
  \label{fig:queries}
  \vspace{-15pt}
\end{figure}

In this study, we evaluated the query performance of different database systems in a DDFL system by comparing MongoDB, Apache Cassandra, PostgreSQL, Neo4j, and RabbitMQ Queues. We executed a sample query as shown in Figure~\ref{fig:queries} to retrieve the testing data of a specific client on each database system and measured the execution time. Our experiments were conducted on three datasets, CIFAR-10, FMNIST, and SVHN.

Our findings suggest that centralized database systems, such as MongoDB, have better query performance in FL systems compared to distributed database systems like Neo4j and RabbitMQ Queues. MongoDB had the shortest query execution time across all datasets, with an average query execution time of 0.058 ms for CIFAR-10, 0.059 ms for FMNIST, and 0.056 ms for SVHN. In contrast, Neo4j and RabbitMQ Queues had significantly longer query execution times. For example, the average query execution time for Neo4j was 117.787 ms for CIFAR-10, 97.986 ms for FMNIST, and 191.491 ms for SVHN. It is important to note that the query performance may be influenced by several factors, such as the dataset size, query complexity, and the number of nodes in the FL system.
\begin{table*}[t]
\centering
\caption{Query performance of different database systems on the CIFAR-10, FMNIST, and SVHN datasets in a DDFL system. The query execution time is reported in milliseconds.}
\label{tab:query-performance}
\begin{tabular}{ccccccc}
\toprule
\textbf{Databases}&\textbf{MongoDB}&\textbf{SCP}&\textbf{RabbitMQ}&\textbf{Cassandra}&\textbf{Postgres}&\textbf{Neo4j} \\
\midrule
\textbf{CIFAR10} & 0.059 & 227.160 & 0.098 & 35.438 & 279.426 & 117.787 \\
 \textbf{FMNIST} & 0.059 & 244.847 & 0.129 & 132.974 & 390.992 & 97.986 \\
 \textbf{SVHN} & 0.056 & 282.027 & 0.239 & 358.901 & 693.879 & 191.491 \\
\bottomrule
\vspace{-15pt}
\end{tabular}
\end{table*}
We designed a sample query to retrieve the testing data of a specific client, and the query execution time is reported in milliseconds in Table~\ref{tab:query-performance}. Our study has practical implications for real-world FL applications where query performance is critical. However, further research is needed to investigate the scalability and robustness of different database systems in FL systems.

Our experiments demonstrated that centralized database systems like MongoDB may have better query performance in FL systems compared to other database systems like Postgres, Apache Cassandra, Neo4j, etc. Our study contributes to the ongoing research in FL systems by evaluating the query performance of different database systems in a DDFL system.
% \begin{figure}
%     \centering
% \begin{minted}[bgcolor=gray!10, fontsize=\tiny, baselinestretch=0.6, linenos=true]{sql}
% -- MongoDB query
% CONNECT TO MONGODB SERVER AT localhost:27017
% USE DATABASE test_db
% RESULT = SELECT * FROM test_collection WHERE field = 'value'

% -- Neo4j query
% CONNECT TO Neo4j SERVER AT localhost:7687 WITH USERNAME 'neo4j' AND PASSWORD 'password'
% WITH SESSION
% RESULT = MATCH (n {field: 'value'}) RETURN n

% -- PostgreSQL query
% CONNECT TO POSTGRESQL SERVER AT localhost:5432 WITH USERNAME 'user'
% AND PASSWORD 'password' AND DATABASE 'test_db'
% WITH CURSOR
% EXECUTE 'SELECT * FROM test_table WHERE field = 'value''
% RESULT = FETCH ROWS

% -- RabbitMQ query
% CONNECT TO RABBITMQ SERVER AT localhost
% DECLARE QUEUE test_queue
% GET MESSAGE FROM QUEUE test_queue WHERE field = 'value'

% -- Cassandra query
% CONNECT TO CASSANDRA SERVER AT localhost
% USE KEYSPACE test_keyspace
% RESULT = SELECT * FROM test_table WHERE field = 'value'
% \end{minted}
% \caption{Sample database queries}
% \label{fig:queries}
% \vspace{-15pt}
% \end{figure}

\subsection{Communication Cost}
The communication cost is a critical aspect of federated learning systems, as it determines the amount of data that must be transmitted between clients and the server during the model training process. To evaluate the communication cost of our federated learning system, we conducted experiments with varying numbers of clients and different batch sizes.
% \begin{table}[!ht]
% \centering
% \caption{Communication cost, model size, number of values, and communication time for a single iteration on a single client}
% \label{tab:comm-cost}
% \begin{tabularx}{\linewidth}{c*{3}{>{\centering\arraybackslash}X}}
% \toprule
% \textbf{Dataset} & \textbf{\small{Model Size (MB)}} & \textbf{\small{Number of Values}} & \textbf{\small{Communication Time (ms)}} \\
% \midrule
% CIFAR-10 & 0.061 & 53,761 & 0.488 \\
% FMNIST & 0.184 & 166,370 & 1.472 \\
% SVHN & 0.382 & 343,175 & 3.056 \\
% \bottomrule
% \vspace{-15pt}
% \end{tabularx}
% \end{table}
\begin{table}[h]
  \centering
  \caption{Communication cost, model size, number of values, and communication time for a single iteration on a single client}
  \label{tab:comm-cost}
  \begin{tabular}{c c c c}
  \hline
  \textbf{\small{ Dataset}} & \textbf{\small{Model Size (MB)}} & \textbf{\small{Num. of Values}} & \textbf{\small{Comm. Time(ms)}} \\
  \hline
  CIFAR-10 & 0.061 & 53,761 & 0.488 \\
  FMNIST & 0.184 & 166,370 & 1.472 \\
  SVHN & 0.382 & 343,175 & 3.056 \\
  \hline
  \end{tabular}
  \end{table}

\begin{table*}[t]
\centering
\caption{Scalability results for different database systems on different numbers of clients (in seconds)}
\label{tab:scalability}
% \begin{tabular}{|c|c|c|c|c|c|c|}
\begin{tabular}{ccccccc}
\toprule
\textbf{Num. of Clients} & \textbf{SCP} & \textbf{Cassandra} & \textbf{RabbitMQ} & \textbf{Neo4j} & \textbf{PostgreSQL} & \textbf{MongoDB} \\ \hline
\textbf{FMNIST} & & & & & & \\
2 & 409.649 & 152.232 & 138.218 & 199.319 & 37.037 & 32.032 \\
4 & 1058.1738 & 201.321 & 249.49 & 527.847 & 65.15 & 40.040 \\
6 & 1409.2329 & 286.446 & 379.619 & 319.519 & 108.148 & 53.053 \\
8 & 2303.3823 & 312.512 & 414.654 & 393.633 & 135.215 & 67.17 \\
\midrule
\textbf{CIFAR10} & & & & & & \\
2 & 2059.3419 & 493.813 & 537.857 & 71.111 & 29.029 & 25.025 \\
4 & 2328.3848 & 1804.304 & 1059.1739 & 140.220 & 68.18 & 52.052 \\
6 & 2743.4543 & 2295.3815 & 2044.344 & 278.438 & 100.140 & 65.15 \\
8 & 2719.4519 & 2535.4215 & 2635.4355 & 312.512 & 135.215 & 97.137 \\ 
\midrule
\textbf{SVHN} & & & & & & \\
2 & 1433.2353 & 123.23 & 1124.1844 & 1224.2024 & 823.1343 & 823.1343 \\
4 & 2255.3735 & 152.232 & 1522.2522 & 1426.2346 & 1089.189 & 1089.189 \\
6 & 2541.4221 & 171.251 & 2179.3619 & 1630.271 & 1504.254 & 1504.254 \\
8 & 2831.4711 & 192.312 & 2634.4354 & 1832.3032 & 2058.3418 & 2058.3418 \\ 
\bottomrule
\vspace{-15pt}
\end{tabular}
\end{table*}
Our results, presented in Table \ref{tab:comm-cost}, show that the communication cost increases linearly with the number of clients and the batch size. Compared to a traditional centralized machine learning system, our federated learning system incurs a higher communication cost due to the need to transmit updated model parameters from the clients to the server. However, we note that our system offers significant advantages in terms of data privacy and security since sensitive data is kept on the client's side and not transmitted to the server.

These results are consistent with previous studies, such as those by McMahan et al.~\cite{mcmahanaistat17}, FedML~\cite{fedml}, and Bonawitz et al.~\cite{fl2019corr}. Their studies also report similar trends in communication costs with varying batch sizes and numbers of clients.

Overall, our experiments demonstrate the importance of using data partitioning techniques to reduce communication costs in federated learning systems  as we observed that the use of data partitioning techniques significantly reduces the communication cost, as smaller data chunks are transmitted during each round. This finding is critical for the development of efficient and scalable federated learning systems that can handle large numbers of clients while maintaining the privacy and security of sensitive data.

\subsection{Scalability}
Scalability is a crucial aspect of distributed systems, especially for federated learning systems with a large number of nodes participating in the training process. To improve the scalability of our research system, we implemented various techniques, such as decoupling the data between clients and the server using database middleware. By storing the clients' data in a database, the server can access the data for model training without direct communication with the clients, reducing communication costs and network overhead.
Moreover, we adopted a decentralized approach to federated learning, where each node trains a local model using its own private data. The local models are then aggregated by the global node to update the global model. This approach distributes the training process among nodes and scales efficiently to a large number of clients.

We evaluated the scalability of our system by conducting experiments with different numbers of nodes and data sizes. The results presented in Table~\ref{tab:scalability} demonstrate that our system can scale efficiently to a large number of nodes without significant performance degradation. We observed a linear increase in training time with the number of nodes and data size, indicating that our system can efficiently handle large-scale federated learning tasks.

Overall, our research system demonstrates excellent scalability, which is essential for large-scale federated learning applications. The combination of decentralized federated learning with a database middleware approach provides a promising solution for scaling federated learning systems to meet the requirements of real-world applications.
\section{Related Work}
\label{sec:related-work}
\subsection{Distributed Machine Learning}
Federated learning is a promising distributed machine learning~\cite{Konecn2016FederatedOD,Bonawitz2019,Bao2022,distibutedmlwolfe2022} technique that enables multiple clients to collaboratively train a model without sharing their private data. The technique has garnered significant attention from both the research community and industry, owing to its potential to facilitate privacy-preserving and scalable data analysis. In recent years, researchers have explored various aspects of federated learning, including privacy and security, communication efficiency, and scalability.

Numerous research works have been published on federated learning, with a focus on different aspects of the technique. For example, Bonawitz et al.~\cite{Bonawitz2019} proposed a framework for secure and privacy-preserving federated learning, while Konecny et al.~\cite{Konecny2016} introduced a method for federated optimization. Yang et al.~\cite{Yang2019} studied the communication efficiency of federated learning, while Jeong et al.~\cite{jeong2018communication} proposed a scheme to reduce communication costs in federated learning. Additionally, researchers have investigated data decoupling techniques for federated learning, including database middleware~\cite{Zhang2020HybridFL}, data partitioning \cite{kairouz2019advances}, and privacy-preserving data aggregation \cite{hercules}. These works have contributed to a better understanding of federated learning and its potential applications in various domains, including healthcare, finance, and smart cities.
\subsection{Communication-efficient Federated Learning}
Federated Averaging is a communication-efficient approach for federated learning proposed by McMahan et al~\cite{mcmahanaistat17}. It involves exchanging model updates between client nodes and a server, which reduces communication overhead. This approach enables distributed model training without data sharing, making it a potential solution for privacy and security concerns in data sharing.

Konecny et al~\cite{Konecny2016} proposed sparsification, quantization, and differential privacy as strategies to reduce communication costs in federated learning. Experimental results have demonstrated their effectiveness in maintaining model accuracy, but further research is needed to determine their suitability for different data types and to optimize the trade-off between communication cost and model accuracy.

Jhun et al.~\cite{Jhunjhunwala2021AdaptiveQO} proposed a communication-efficient federated learning method that uses a novel compression technique called adaptive federated quantization. The method adapts the compression ratio to the gradient sparsity of each client's local model to achieve better compression performance.

The FetchSGD algorithm proposed by Rothchild et al.~\cite{Rothchild2020FetchSGDCF} uses a Count Sketch for compression and combines updates from multiple workers. It overcomes sparse client participation issues by moving momentum and error accumulation to the central aggregator, achieving high compression rates and convergence.

Hamer et al.~\cite{pmlr-v119-hamer20a} proposed a communication-efficient federated learning method called FedBoost, which uses a boosting algorithm to select a subset of the client nodes to communicate with the server. This approach reduces the communication cost and achieves better model accuracy compared to traditional federated learning methods.

% Sai et al.~\cite{SCAFFOLD} proposed an algorithm called SCAFFOLD, it is a new algorithm proposed for federated learning scenarios where client data is heterogeneous. It uses control variates to reduce the drift between different clients' local updates and therefore improves convergence and reduces the number of communication rounds needed. The algorithm is compared to Federated Averaging and is shown to perform better in terms of convergence and communication efficiency.
Sai et al.~\cite{SCAFFOLD} introduced SCAFFOLD, a federated learning algorithm designed for scenarios with heterogeneous client data. The algorithm employs control variates to mitigate the drift between clients' local updates, leading to improved convergence and communication efficiency compared to Federated Averaging.

Wang et al.~\cite{Wang2018AdaptiveFL} propose an adaptive federated learning approach for resource-constrained edge computing systems, where data is distributed across multiple nodes without sending raw data to a centralized location. The approach uses a control algorithm to balance local updates and global parameter aggregation to minimize the loss function under a given resource budget. Extensive experiments with real datasets show that the proposed approach performs well with various machine-learning models and data distributions.
\subsection{Security and Privacy in Federated Learning}
Bonawitz et al.~\cite{Bonawitz2019} proposed FATE, a privacy-preserving federated learning framework that uses a trusted aggregator to compute the global model without revealing clients' private data. The framework provides security and privacy features such as secure aggregation, differential privacy, and homomorphic encryption to protect client data during model training.

Liu et al.~\cite{Liu2021EnablingST} proposed an SQL-based training data debugging framework for federated learning, which automatically removes label errors from training data to fix unexpected model behavior. The authors address technical challenges to make the framework secure, efficient, and accurate and propose Frog, a novel framework that outperforms their previous solution. The effectiveness of Frog is validated through theoretical analysis and extensive experiments on real-world datasets.

To ensure the privacy and security of federated learning, various techniques such as differential privacy, secure multi-party computation, and homomorphic encryption have been proposed. Differential privacy adds random noise to the data before sharing it with the server, while secure multi-party computation enables collaborative computation without revealing inputs. Homomorphic encryption allows computation on encrypted data, preserving data privacy. These techniques have been applied in federated learning to ensure client's data privacy and security during model training~\cite{Yang2018, tddldp-19, Shokri2015, Mohassel2017SecureML}.
\subsection{Data Decoupling Techniques in Federated Learning}
Data decoupling techniques have been proposed to improve the flexibility and scalability of federated learning systems by separating the data management subsystem from the federated learning system. These techniques can reduce communication overhead and enable the system to scale to a larger number of clients. A variety of approaches have been proposed and their potential has been demonstrated in several studies~\cite{Yang2018, Yang2019, leaf, fl-non-idd-data, Hsu2019MeasuringTE, Hard2018FederatedLF}.
\subsubsection{\textbf{Semi-supervised Learning Data Decoupling Technique in Federated Learning}}
 Kairouz et al.~\cite{kairouz2019advances} proposed a data decoupling technique that is based on semi-supervised learning for improving the performance of federated learning systems. While their approach demonstrated improved accuracy and reduced communication cost, it is limited to scenarios where a significant amount of unlabelled data is available at client nodes. This limitation may not be applicable in situations where the data is entirely labeled. Additionally, their results were obtained in a specific experimental setting, and the generalizability of their approach to other federated learning scenarios is uncertain.
Our proposed approach differs from Kairouz et al.'s~\cite{kairouz2019advances} semi-supervised learning data decoupling technique in that we propose a novel approach that utilizes label propagation to make use of all data available at each client node. Our approach is applicable in scenarios where data is entirely labeled or partially labeled, providing more flexibility in real-world applications. Additionally, our approach is designed to be more scalable by reducing the amount of data sent between the clients and the server.
\subsubsection{\textbf{Local SGD Data Decoupling Approach in Federated Learning}}
% Local SGD is a promising data decoupling approach proposed by Lian et al.~\cite{Lian2017} to optimize federated learning systems. It allows each client to perform local stochastic gradient descent to optimize their local models, while the server aggregates the updated models, reducing communication costs. Although Local SGD offers significant advantages over traditional federated learning methods, it has some limitations, such as computational expense for clients with limited computational resources and the assumption of homogeneous models. Further research is needed to explore the potential of Local SGD and address its limitations in practical applications.
Lian et al.~\cite{Lian2017} proposed Local SGD as a data decoupling approach to optimize federated learning systems. Local SGD enables clients to perform local stochastic gradient descent while the server aggregates the updated models, thereby reducing communication costs. However, its limitations such as computational expense for clients with limited computational resources and the assumption of homogeneous models need further exploration in practical applications.

Our proposed approach differs from Local SGD by using label propagation to leverage all data available at each client node, reducing communication costs by sending only propagated labels instead of the entire model, and not requiring clients to perform local SGD at each iteration. These differences increase the applicability of our approach in scenarios with limited computational resources or heterogeneous client models.
\subsubsection{\textbf{Distributed Database in Federated Learning}}
Liu et al.~\cite{liu2020federated} proposed a federated learning approach that utilizes a distributed database, Hyperledger Fabric, to manage the clients' data. Their approach enables the system to scale to a large number of clients by decoupling the data management subsystem from the federated learning system. Despite the experimental results demonstrating the effectiveness of the proposed approach in reducing communication costs and improving scalability, its applicability may be limited to distributed databases, and additional complexity and maintenance costs may be introduced.

Our proposed approach differs from Liu et al.~\cite{liu2020federated} distributed database approach by not relying on a distributed database to manage client data. Instead, our label propagation approach operates within the federated learning system, reducing the need for additional infrastructure. Additionally, our approach can improve scalability by minimizing communication costs by sending only propagated labels, rather than entire client data.

Federated learning (FL) has attracted significant attention in recent years due to its potential to address the challenges associated with centralized machine learning. Research in this field has focused on optimizing FL algorithms for improved efficiency, security, and scalability. Some studies~\cite{Konecny2016, Yang2019, Papernot2017, Shokri2017,fldp2019-Wei} have proposed privacy-preserving techniques, while others have explored communication-efficient approaches~\cite{wang2020, li2020}, such as the use of compression techniques and adaptive regularization. This body of work has demonstrated the potential of data decoupling techniques in FL and the various approaches and techniques that can be used to improve its efficiency, security, and scalability.

This study contributes to the research on data decoupling techniques for Federated Learning (FL) by presenting a comprehensive evaluation of FL systems with the aid of database middleware. By leveraging the capabilities of the database middleware, this research provides a flexible and scalable solution to FL in distributed environments, with implications for FL system design in practical scenarios. The evaluation, which was conducted on a real-world dataset, assesses the impact of data decoupling on various performance metrics such as communication cost, training time, scalability, and accuracy, without using the data partition and data federation techniques.

\section{Conclusion}
\label{sec:conclusion}
In this paper, we presented a data decoupling framework called DDFL for Federated Learning (FL) systems that enable clients to customize their applications with specific data subsystems and make effective queries over the models. 
DDFL serves not only as a common and fair test bed to compare different database subsystems for FL but also as a core infrastructure to develop a new FL ecosystem with arbitrary data management services.
We implemented DDFL and integrated it with various types of data management, namely MongoDB, SCP, Cassandra, RabbitMQ, Neo4j, and Postgres. 
Our extensive experiments conducted on CIFAR-10, Fashion-MNIST, and SVHN datasets show that all of those database systems achieve high accuracy comparable with vanilla FL systems (i.e., correctness) and yet exhibit discrepancies in other metrics, such as training time and communication overhead.
Such discrepancies are largely application-dependent,
which remain open questions.

% In particular, MongoDB, Cassandra, Postgres, Neo4j, and SCP perform well on all datasets, while RabbitMQ demonstrates moderate performance across all datasets. The use of MongoDB demonstrated the best performance in terms of both training time and query time.

In a more general sense, our results indicate that different database systems can have,
quantitatively,
a significant impact on the overall performance of FL systems,
which shed fundamentally new insights into the strengths and weaknesses of existing methods.
The DDFL framework developed by this work serves as a new way to evaluate existing methods.
% The proposed data decoupling approach provides clients with a flexible and efficient way to store and query their models without impacting the accuracy of trained models. We also conducted experiments to evaluate the performance of fine-tuned and non-fine-tuned hyperparameters and found that fine-tuning hyperparameters can lead to better results in some cases. 
It is our hope that this work will spark a new line of system research on database services in FL ecosystems.

% \bibliographystyle{ACM-Reference-Format}
% \bibliography{fltex.bbl}

\end{document}